# Thermoelectric properties of InAs nanowires from full-band atomistic simulations

Damiano Archetti and Neophytos Neophytou*

School of Engineering, University of Warwick, Coventry, CV4 7AL, UK

*N.Neophytou@warwick.ac.uk

## Abstract

In this work we theoretically explore the effect of dimensionality on the thermoelectric power factor of InAs nanowires by coupling atomistic tight-binding calculations to the Linearized Boltzmann transport formalism. We consider nanowires with diameters from 40nm (bulk-like) down to 3nm (1D), which allows for the proper exploration of the power factor within a unified large-scale atomistic description across a large diameter range. We find that as the diameter of the nanowires is reduced below $d <$ 10 nm, the Seebeck coefficient increases substantially, a consequence of strong subband quantization. Under phonon-limited scattering conditions, a considerable improvement of ~6× in the power factor is observed around $d$ = 10 nm. The introduction of surface roughness scattering in the calculation reduces this power factor improvement to ~2×. As the diameter is decreased down to $d$ = 3 nm, the power factor is diminished. Our results show that, although low effective mass materials such as InAs can reach low-dimensional behavior at larger diameters and demonstrate significant thermoelectric power factor improvements, surface roughness is also stronger at larger diameters, which takes most of the anticipated power factor advantages away. However, the power factor improvement that can be observed around $d$ = 10 nm, could prove to be beneficial as both the Lorenz number and the phonon thermal conductivity are reduced at that diameter. Thus, this work, by using large-scale full-band simulations that span the corresponding length scales, clarifies properly the reasons behind power factor improvements (or degradations) in low-dimensional materials. The elaborate computational method presented can serve as a platform to develop similar schemes for 2D and 3D material electronic structures.

# I. Introduction

The efficiency of thermoelectric (TE) materials is quantified by the dimensionless figure of merit $ZT$ as:

$$ZT = \frac{\sigma S^2}{\kappa_e + \kappa_l} \quad (1)$$

where $\sigma$ is the electrical conductivity, $S$ is the Seebeck coefficient, and $\kappa$ is the thermal conductivity composed of two parts, the electronic part of the thermal conductivity $\kappa_e$, and the phonon/lattice part of the thermal conductivity $\kappa_l$. The quantity $\sigma S^2$ is the power factor (PF). Over the last several years, a myriad of materials and concepts for high $ZT$ have evolved [1] including GeTe [2], PbTe [3], half-Heuslers [4], skutterudites [5], etc.. Low dimensional materials such as nanowires (NWs) are one of these, as they can achieve extremely low thermal conductivities due to strong phonon-interface scattering. Significant increases in TE performance and $ZT$ in NWs and their networks have been reported [6, 7, 8, 9, 10, 11, 12, 13]. $ZT$ values up to 1 for NWs based on several materials (Si, SiGe, InAs, InSb, Bi, PbTe, ZnO, SnSe, NiFe, and many more) have been investigated [11, 14, 15, 16]. Since the pioneering work by Hicks and Dresselhaus, efforts have also been focused on utilizing the sharp features in the low-dimensional density-of-states to improve the power factor as well [17, 18]. Theoretical studies on the thermoelectric power factor of NWs, showed that 1D modes could provide power factor improvements even up to 30% [19, 20, 21, 22]. Experimentally, however, this has not yet been achieved, because to observe the true 1D nature, one needs to consider NW diameters down to a few nanometers (as in the case of Si) [23]. At those dimensions, however, surface roughness scattering (SRS) drastically reduces $\sigma$ [23], but also distorts the sharp features in the density-of-states [24].

Low-dimensional effects could be evident at larger length scales in low effective mass materials such as InAs, InSb, or Bi [25]. Utilizing these materials could be technologically more feasible, but $\sigma$ could also be less susceptible to surface roughness scattering (SRS) in thicker channels. For example, strong subband quantization and bandstructure effects begin to appear in the case of Si NW channels (with effective mass $m^* \sim 0.2\ m_0$,) at diameters below ~10nm [23]. In the case of InAs, however, with effective mass $m^*$ ~ 0.02 $m_0$, we expect such effects to appear at a larger length scale, as it was observed in the case of Bi nanowires as well [25, 26]. The Seebeck coefficient, in particular, as we have previously shown, begins to increase in an almost linear fashion with diameter reduction from the point where quantum confinement splits the NW subbands at such degree, which leaves only a few subbands (ideally one) in the vicinity of the Fermi level [19, 23]. Thus, low-effective mass materials, which reach the 'few subband' condition at larger diameters, could provide a larger Seebeck coefficient increase with further diameter reduction compared to channels with larger effective masses. Nanowires with larger diameters are practically more feasible and controllable as well. Power factor benefits would then be more easily realized. The subband quantization, a signature of low dimensionality, has been observed at lower temperatures in InAs NWs, where the effect of individual subband features was observed in all three coefficients, the electrical conductivity, the Seebeck coefficient and the power factor [27, 28]. In another low temperature work, InAs/InP NW superlattices were fabricated, and quantum dots were formed, exhibiting promising thermoelectric energy power extraction and conversion efficiency [29]. Doping and planar defects are also investigated in order to optimize the PF and decrease the NW thermal conductivity [30]. References [31, 32], have also measured promising TE performance for InAs NWs with diameters as low as 20 nm.

In this work we explore the thermoelectric properties of InAs nanowires with a focus on the effects of dimensionality on the power factor. Our intent it is provide general understanding on the topic as well. We describe a computational method which couples large-scale atomistic tight-binding electronic structures with analytical wavefunction descriptions to Boltzmann transport with energy-dependent scattering times, going beyond the commonly employed constant relaxation time approximation. We consider NWs with diameters from 40 nm down to 3 nm (calculations including up to 30,000 atoms). We show that in InAs NWs, low-dimensionality effects begin to influence the bandstructure and transport at diameters *d* ~ 20 nm. Significant Seebeck coefficient, surprisingly electrical conductivity, and power factor improvements compared to bulk InAs are observed as the diameter is scaled in the phonon-limited transport case, even up to 6× for the power factor of the *d* ~ 10 nm NW. The introduction of SRS reduces the conductivity and PF. Still, however, we find that for NWs with diameters around *d* ~ 10 nm, a ~2× PF improvement is retained, and bulk or higher PF values are observed down to $d \sim 7$ nm. This could prove promising because at such diameters the phonon thermal conductivity $\kappa_l$ is reduced and $ZT$ can be improved [6, 7, 33]. We believe that our results will further add to the understanding of the effects of low-dimensionality on the PF and the conditions under which improvements can be observed. We also believe that the method we employ can prove useful in other TE material investigations which require capturing accurately the energy dependence of the scattering times, especially when extended to 2D and 3D materials.

The paper is organized as follows: In Section II we describe our theoretical and computational approach. In Section III we present the thermoelectric properties of the InAs nanowires under investigation and discuss the results. In Section IV we conclude.

## II. Theoretical and computational method

The electronic structures of the nanowires are calculated using the $sp^3d^5s^*$ tight-binding model with the parametrization of Ref. [34]. The model is validated to capture all of the relevant features of the bandstructure of semiconductors that appear at the nanoscale. Previous works showed that tight-binding methods could capture essential bandstructure features beyond band quantization, such as band splitting, non-parabolicity, band warping, effective mass variation, etc. [35, 36, 37, 38, 39, 40, 41]. Importantly for this work, tight-binding is robust enough to calculate the bandstructure for NWs up to 40 nm in diameter that we consider (structures of up to 30,000 atoms in the unit cell). We consider [100] n-type InAs nanowires, in which case we ignore spin-orbit coupling. The bandstructures of InAs NWs of diameters $d$ = 3 nm, 10 nm, 20 nm, and 40 nm are shown in Fig. 1a-d. The position of the Fermi level for the carrier density of $n = 10^{18}/\text{cm}^3$ (approximate concentration where the thermoelectric power factor peaks for the $d$ = 10 nm NW) is indicated by the red lines in each sub-figure. The two important things to notice here as the diameter is scaled are the following: i) the number of subbands is reduced to very few, even to a single subband, and ii) the position of the Fermi level, which directly determines the Seebeck coefficient ($S$), shifts lower compared to the band edge (comparing here at the same carrier density). Indeed, the distance of the Fermi level from the band edge $\eta_F = E_C - E_F$ increases substantially with diameter reduction as shown in the inset of Fig. 1d (in units of $k_B T$). The Seebeck coefficient is proportional to the average energy of the current flow as $S \sim \langle E-E_F \rangle$, which depends linearly on $\eta_F$. This increase in $\eta_F$ originates from the fact that the number of subbands decreases slower compared to the NW area (and cannot be reduced to zero subbands at ultra-narrow diameters). The only way then to retain a constant carrier (3D) density is to increase $\eta_F$ (to lower $E_F$), and this is the reason that Seebeck coefficient improvements are expected at low dimension as we show below [19, 23].

Bandstructure features under confinement: Two other important features of the InAs bandstructure that affect the transport properties are the variations of the effective mass and the differential of the shift in the band edges under diameter scaling. These are shown in Fig. 2a and 2b, respectively. The effective mass of a channel determines its mobility by a large amount as $\mu \sim m^{*(-3/2)}$, whereas the differential of the band edges determines the strength of the SRS rate as $r_{SRS} \sim (\Delta E/\Delta d)^2$ [42]. Results for the two nanowire orientations [100] (blue lines) and [110] (orange lines) are shown. Both quantities begin to increase when the diameter is scaled below $d \sim 20$ nm, whereas significant increases are observed for diameters below $d \sim 10$ nm. The increase in the effective mass originates from the behavior of non-parabolic bands under confinement and is well explained in previous works and was observed in Si [35, 36] and Bi nanowires as well [43]. For diameters down to $d = 3$ nm an increase of ~3× is observed. In contrast, in the case of Si for example, a less non-parabolic material, the corresponding increase in the effective masses is somewhat less than half up to 1.4× (inset of Fig. 2a). In a similar way, the differential of the shifts in the band edges increase largely with diameter reduction (Fig. 2b), in a more sensitive fashion compared to Si channels (inset of Fig. 2b). Both the effective masses and band edges are more sensitive to quantization in InAs compared to Si due to the much smaller effective mass. This sensitivity, however, would have negative impact to its transport properties, as we show below.

Transport – Linearized BTE formalism: In order to obtain the TE coefficients we employ the Linearized Boltzmann Transport Equation (BTE) as described in [23, 45, 46]. Within the BTE, the TE coefficients are given by:

$$\sigma = e_0^2 \int_{E_C}^{\infty} dE \left( -\frac{\partial f_0}{\partial E} \right) \Xi(E), \tag{2}$$

$$S = \frac{e_0 k_B}{\sigma} \int_{E_C}^{\infty} dE \left( -\frac{\partial f_0}{\partial E} \right) \Xi(E) \left( \frac{E - \mu_1}{k_B T} \right), \tag{3}$$

$$\kappa_0 = k_B{}^2 T \int_{E_C}^{\infty} dE \left( -\frac{\partial f_0}{\partial E} \right) \Xi(E) \left( \frac{E - \mu_1}{k_B T} \right)^2, \tag{4}$$

$$\kappa_e = \kappa_0 - T\sigma S^2 \tag{5}$$

where $e_0$ is the charge of the electron, $\frac{\partial f_0}{\partial E}$ is the derivative of the Fermi distribution, and $\Xi(E)$ is the transport distribution function, defined in the multiband approach that we employ as:

$$\begin{aligned} \Xi(E) &= \frac{1}{A} \sum_{k_x, n} v_n{}^2(k_x) \tau_n(k_x) \delta(E - E_n(k_x)) \\ &= \frac{1}{A} \sum_{k_x, n} v_{k_x, n}{}^2(E) \tau_{k_x, n}(E) g_{k_x, n}^{1D}(E) \end{aligned} \tag{6}$$

where $n$ is the subband index, $k_x$ is the wavevector index, $\tau_{k_x,n}(E)$ is the carrier relaxation time, $g_{k_x,n}^{1D}(E)$ is the one-dimensional density of states, and $v_k$ is the band velocity computed as:

$$v_k = \frac{1}{\hbar} \frac{\partial E}{\partial k} \tag{7}$$

We perform all simulations this work for room temperature $T$ = 300 K and we consider electron-phonon scattering and surface roughness scattering (SRS). It is important to stress that this approach allows for the full energy/momentum dependence of the relaxation times, which turns to be very important in evaluating TE material properties [47], but routinely omitted due to the computational burden that it involves.

The calculation of the relaxation times depends on the particular scattering mechanism under consideration, and formal treatment of electron-phonon scattering can be found in Ref. [48], and more specifically for NWs within the full-band atomistic description in Ref. [23]. We repeat some of the initial mathematical steps here for completion. Within the Fermi's Golden Rule approach, the scattering rate of an electron from an initial state with wave vector $k_x$ to a final state $k_x'$ interacting with a phonon is calculated as:

$$S_{n,m}(k_x, k_x') = \frac{2\pi}{\hbar}\left|H^{m,n}_{k_x',k_x}\right|^2 \delta\left(E_m(k_x') - E_n(k_x) - \Delta E\right). \tag{8}$$

The relaxation times are then calculated by:

$$\frac{1}{\tau_n(k_x)} = \sum_{m,k_x'} S_{n,m}(k_x, k_x') \tag{9}$$

where the $H^{m,n}_{k_x',k_x}$ and $S_{n,m}(k_x, k_x')$ quantities result after integrating out all radial components.

The matrix element is computed using the scattering potential $U_S(\vec{r})$ as:

$$H^{m,n}_{k_x',k_x} = \int_{-\infty}^{\infty} \psi_{m,k_x'}{}^{*}(\vec{r}) U_S(\vec{r}) \psi_{n,k_x}(\vec{r}) d^3r \tag{10}$$

The matrix element then becomes:

$$H^{m,n}_{k_x',k_x} = \frac{1}{\Omega}\int_{-\infty}^{\infty} F_m{}^{*}(\vec{R}) e^{-ik_x'x} U_S(\vec{r}) F_n(\vec{R}) e^{ik_x x} d^2R dx. \tag{11}$$

We consider three dimensional phonons with total wavevector $q = q_{\mathrm{R}} + q_{\mathrm{x}}$, which can be decomposed in a component $q_{\mathrm{x}}$ parallel to the transport direction (assumed in *x*-direction) and a component $q_{\mathrm{R}}$ that lies in the cross section radial plane. The perturbing potential in the case of phonon scattering is then defined as:

$$U_S(\vec{r}) = A_{\vec{q}} K_{\vec{q}} e^{\pm i(\vec{q}\cdot\vec{r} - \omega t)} \tag{12}$$

where $A_{\vec{q}}$ is associated with the lattice vibration amplitude and $K_{\vec{q}}$ with the deformation potential. Above $\left|A_{\vec{q}}\right|^2 = \frac{1}{\Omega}\frac{\hbar\left(N_\omega + \frac{1}{2} \mp \frac{1}{2}\right)}{2\rho\omega_{\mathrm{ph}}}$, where $\rho$ is the mass density, $N_\omega$ is the number of phonons given by the Bose-Einstein distribution, and $\Omega$ is the unit cell volume. For acoustic deformation potential scattering (ADP), optical deformation potential scattering (ODP), and for polar optical phonon scattering (POP), respectively, for the strength of the scattering event, it holds:

$$\left|K_{\vec{q}}\right|^2 = q^2 {D_{\mathrm{ADP}}}^2 \text{ for ADP,} \tag{13a}$$

$$\left|K_{\vec{q}}\right|^2 = {D_{\mathrm{O}}}^2 \text{ for ODP,} \tag{13b}$$

$$\left|K_{\vec{q}}\right|^2 = \frac{\rho {e_0}^2 {\omega_{\mathrm{ph}}}^2}{\left({q_{\mathrm{R}}}^2 + {q_{\mathrm{x}}}^2\right)}\left(\frac{1}{\varepsilon_\infty} - \frac{1}{\varepsilon_{\mathrm{s}}}\right) \text{ for POP} \tag{13c}$$

where $D_{\mathrm{ADP}}$ and $D_{\mathrm{O}}$ are the scattering deformation potential amplitudes, $\varepsilon_\infty$ is the high frequency dielectric constant and $\varepsilon_{\mathrm{s}}$ is the static dielectric constant (for InAs we use $\varepsilon_\infty = 12.3$, and $\varepsilon_{\mathrm{S}} = 15.15$) [49, 50].

The scalar product of the phonon wavevector can be split as $\vec{q}\cdot\vec{r} = \vec{q}_{\mathrm{R}}\cdot\vec{R} + q_{\mathrm{x}} x$, and in this case the matrix element becomes:

$$H_{k_x',k_x}^{m,n} = \int_{-\infty}^{\infty} \frac{{F_m}^*\left(\vec{R}\right)}{\sqrt{A}} \frac{e^{-ik_x' x}}{\sqrt{L_x}} A_{\vec{q}} K_{\vec{q}} e^{\pm i\vec{q}_{\mathrm{R}}\cdot\vec{R}} e^{\pm i q_x x} \frac{F_n\left(\vec{R}\right)}{\sqrt{A}} \frac{e^{ik_x x}}{\sqrt{L_x}} d^2R dx, \tag{14}$$

where $L_{\mathrm{x}}$ is the length of the unit cell, $F_{n/m}\left(\vec{R}\right)$ is the cross sectional part of the wave function of the initial/final state, and the integral is performed over the cross section of the nanowire.

Since no other $x$-dependent quantities are found other than in the two exponentials (not in $A_{\vec{q}}$ and not in $K_{\vec{q}}$), the integral over the transport $x$-direction becomes a Kronecker-delta expressing momentum conservation in the transport direction:

$$H_{k_x',k_x}^{m,n} = I_{k_x',k_x}^{m,n}\left(\vec{q}_{\mathrm{R}}\right) A_{\vec{q}} K_{\vec{q}} \delta_{k_x',k_x \pm q_x} \tag{15}$$

with the wavefunction overlap form factor being:

$$I_{k_x',k_x}^{m,n}\left(\vec{q}_{\mathrm{R}}\right) = \frac{1}{A}\int_R {F_m}^*\left(\vec{R}\right) F_n\left(\vec{R}\right) e^{\pm i\vec{q}_{\mathrm{R}}\cdot\vec{R}} d^2R, \tag{16}$$

Following Ref. [51] we assume simple envelope wave functions which are constant inside the nanowire and zero outside as:

$$\frac{F(R)}{\sqrt{A}}=\frac{\Theta(R_0-R)}{\sqrt{A}}=\frac{\Theta(R_0-R)}{\sqrt{\pi}R_0} \tag{17}$$

where $R_0$ is the nanowire radius and $\Theta(\bullet)$ is the Heaviside function.

To obtain the transition rate $S_{n,m}(k_x,k_x')$, we then square the matrix element $H^{m,n}_{k_x',k_x}$ and sum over all lateral momenta $q_\mathrm{R}$ as:

$$S_{n,m}(k_x,k_x')=\frac{2\pi}{\hbar}\sum_{q_\mathrm{R}}\left|I^{m,n}_{k_x',k_x}\right|^2\left|K_{\vec{q}}\right|^2\left|A_{\vec{q}}\right|^2\delta_{k_x',k_x\pm q_x}\delta\left(E_m(k_x')-E_n(k_x)\pm\hbar\omega_{ph}\right) \tag{18}$$

Note that the integral in Eq. (16) is performed only over the confined coordinates since the integral in the transport coordinate is already included in the delta-function for momentum conservation.

By expanding the terms in the summation, transforming the summation into an integral, using the constant wavefunctions of Eq. (17), and the definition of Bessel functions, we reach:

$$\left|D_{\mathrm{1D}}(q)\right|^2=\sum_{q_\mathrm{R}}\left|I^{m,n}_{k_x',k_x}\right|^2\left|K_{\vec{q}}\right|^2\left|A_{\vec{q}}\right|^2 = \frac{2A}{\pi}\left[\int_0^\infty dq_\mathrm{R}\,\frac{J_1^{\,2}(q_\mathrm{R}R_0)}{q_\mathrm{R}R_0^{\,2}}\left|K_{\vec{q}}\right|^2\left|A_{\vec{q}}\right|^2\right] \tag{19}$$

where $J_1(\bullet)$ is the Bessel function of first kind of order 1 (see Appendix for full derivation).

For acoustic deformation potential scattering (ADP) using the usual equipartition approximation $N_\omega \simeq N_\omega+1$, and $N_\omega \simeq k_BT/\hbar\omega_\mathrm{ph}$ because $\hbar\omega_\mathrm{ph}\ll k_BT$ [48], including both emission and absorption processes, and using the fact that $\int_0^\infty dx\frac{J_1^{\,2}(x)}{x}=\frac{1}{2}$, we obtain (see Appendix for full derivation):

$$\left|D_{\mathrm{1D}}(q)\right|^2=\frac{1}{\Omega}\frac{D_{\mathrm{ADP}}^{\,2}k_\mathrm{B}T}{\rho v_\mathrm{s}^{\,2}} \tag{20}$$

where $D_\mathrm{ADP}$ is the acoustic deformation potential, $T$ is the temperature and $v_\mathrm{s}$ is the velocity of sound in the material (for InAs we use $D_\mathrm{ADP}=10eV$ [52, 53, 54, 55, 56], $v_\mathrm{S}=4280\ m/s$, and $\rho=5667\ kg/m^3$ [53]). Inserting this into Eq. (18), and after separating the volume Ω into the cross sectional area $A$, and the longitudinal unit cell length $L_X$, we obtain the transition rates and relaxation times as:

$$S_{n,m}\left(k_x,k_x'\right)=\frac{2\pi}{\hbar}\frac{{D_{\mathrm{ADP}}}^2 k_{\mathrm{B}}T}{A\rho {v_{\mathrm{s}}}^2}\frac{1}{L_{\mathrm{x}}}\delta_{k_x',k_x\pm q_x}\delta\left(E_m\left(k_x'\right)-E_n\left(k_x\right)\right) \tag{21}$$

$$\frac{1}{\tau_n\left(k_x\right)}=\frac{2\pi}{\hbar}\frac{{D_{\mathrm{ADP}}}^2 k_B T}{A\rho {v_{\mathrm{s}}}^2}\left[\frac{1}{L_{\mathrm{x}}}\sum_{m,k_x'}\delta_{k_x',k_x\pm q_x}\delta\left(E_m\left(k_x'\right)-E_n\left(k_x\right)\right)\right] \tag{22}$$

The term in the bracket is simply one half of the one-dimensional density of states, accounting for the fact that the final scattering states have the same spin orientation as the initial state (the convention for the density of states $g_{\mathrm{1D}}\left(E\right)$ contains both spins) [48]. We then replace the summation by integration over energy with the δ-function resulting in the density of final states $g_{\mathrm{1D}}\left(E'\right)$, which in the case of elastic acoustic phonon scattering is the same as the initial density of states $g_{\mathrm{1D}}\left(E\right)$. Thus, we can simplify the scattering rate expression to:

$$\frac{1}{\tau_{n,k_x}^{\mathrm{ADP}}\left(E\right)}=\frac{\pi {D_{\mathrm{ADP}}}^2 k_B T}{\hbar\rho {v_{\mathrm{s}}}^2 A}g_{\mathrm{1D}}\left(E\right) \tag{23}$$

For optical deformation potential scattering (ODP), in a similar way we obtain (see Appendix for full derivation):

$$\left|D_{\mathrm{1D}}\left(q\right)\right|^2=\frac{1}{\Omega}{D_{\mathrm{ODP}}}^2\frac{\hbar\left(N_\omega+\frac{1}{2}\mp\frac{1}{2}\right)}{2\rho\omega_{\mathrm{ph}}} \tag{24}$$

where $D_{\mathrm{ODP}}$ is the optical deformation potential (we use $D_{\mathrm{ODP}}=2\times10^{10}eV/m$) [54]. Inserting this into Eq. (18), and after separating the volume Ω into the cross sectional area $A$, and the longitudinal unit cell length $L_X$, we obtain the transition rates and relaxation times as:

$$S_{n,m}\left(k_x,k_x'\right)=\frac{\pi {D_{\mathrm{ODP}}}^2\left(N_\omega+\frac{1}{2}\mp\frac{1}{2}\right)}{A\rho\omega_{\mathrm{ph}}}\frac{1}{L_{\mathrm{x}}}\delta_{k_x',k_x\pm q_x}\delta\left(E_m\left(k_x'\right)-E_n\left(k_x\right)\pm\hbar\omega_{\mathrm{ph}}\right) \tag{25}$$

$$\frac{1}{\tau_n(k_x)} = \frac{\pi D_{\mathrm{ODP}}{}^2\left(N_\omega + \frac{1}{2} \mp \frac{1}{2}\right)}{A\rho\omega_{ph}} \left[\frac{1}{L_{\mathrm{x}}}\sum_{m,k_x'} \delta_{k_x',k_x \pm q_x} \delta\left(E_m(k_x') - E_n(k_x) \pm \hbar\omega_{ph}\right)\right] \quad (26)$$

As earlier, the term in the bracket above is one half of the one-dimensional density of final states $g_{\mathrm{1D}}(E') = g_{\mathrm{1D}}\left(E \pm \hbar\omega_{ph}\right)$ and the rate can be simplified to:

$$\frac{1}{\tau_{n,k_x}^{\mathrm{ODP}}(E)} = \frac{\pi D_{\mathrm{ODP}}{}^2\left(N_\omega + \frac{1}{2} \mp \frac{1}{2}\right)}{2\rho\omega_{ph}A} g_{\mathrm{1D}}\left(E \pm \hbar\omega_{ph}\right) \quad (27)$$

For polar optical phonon (POP) scattering, the dominant scattering mechanism in polar materials, by exploring Bessel function definitions, we obtain (see Appendix for full derivation):

$$\left|D_{\mathrm{1D}}(q)\right|^2 = e_0{}^2\omega_{\mathrm{ph}}\hbar\left(\frac{1}{\varepsilon_\infty} - \frac{1}{\varepsilon_{\mathrm{s}}}\right)\frac{\left(N_\omega + \frac{1}{2} \mp \frac{1}{2}\right)}{\Omega}\left[\frac{1}{q_{\mathrm{x}}{}^2}\left(\frac{1}{2} - I_1\left(\left|q_{\mathrm{x}}R_0\right|\right)K_1\left(\left|q_{\mathrm{x}}R_0\right|\right)\right)\right] \quad (28)$$

where $I_1(x)$ and $K_1(x)$ are the modified Bessel functions of the first order of the first and second kind, respectively. Inserting this into Eq. (18), and after separating the volume Ω into the cross sectional area *A*, and the longitudinal unit cell length $L_{\mathrm{X}}$, we obtain the transition rates and relaxation times as:

$$S_{n,m}(k_x, k_x') = \frac{\pi}{A} e_0{}^2\omega_{\mathrm{ph}}\left(\frac{1}{\varepsilon_\infty} - \frac{1}{\varepsilon_{\mathrm{s}}}\right)\left(N_\omega + \frac{1}{2} \mp \frac{1}{2}\right)\frac{1}{L_{\mathrm{x}}}\left[\frac{1}{q_{\mathrm{x}}{}^2}\left(1 - 2I_1\left(\left|q_{\mathrm{x}}R_0\right|\right)K_1\left(\left|q_{\mathrm{x}}R_0\right|\right)\right)\right] \delta_{k_x',k_x \pm q_x}\delta\left(E_m(k_x') - E_n(k_x) \pm \hbar\omega_{ph}\right) \quad (29)$$

$$\frac{1}{\tau_n(k_x)} = \frac{\pi}{A} e_0{}^2\omega_{\mathrm{ph}}\left(\frac{1}{\varepsilon_\infty} - \frac{1}{\varepsilon_{\mathrm{s}}}\right)\left(N_\omega + \frac{1}{2} \mp \frac{1}{2}\right)\left[\frac{1}{L_{\mathrm{x}}}\sum_{m,k_x'}\left[\frac{1}{q_{\mathrm{x}}{}^2}\left(1 - 2I_1\left(\left|q_{\mathrm{x}}R_0\right|\right)K_1\left(\left|q_{\mathrm{x}}R_0\right|\right)\right)\right] \delta_{k_x',k_x \pm q_x}\delta\left(E_m(k_x') - E_n(k_x) \pm \hbar\omega_{ph}\right)\right] \quad (30)$$

As earlier, the energy δ-function picks the one-dimensional density of final states $g_{\mathrm{1D}}(E') = g_{\mathrm{1D}}\left(E \pm \hbar\omega_{ph}\right)$ (half of it to account for same spin of initial and final states) and the rate can be simplified to:

$$\frac{1}{\tau_{n,k_x}^{\mathrm{POP}}(E)} = \frac{\pi e_0{}^2 \omega_{\mathrm{ph}}}{2A}\left(\frac{1}{\varepsilon_\infty} - \frac{1}{\varepsilon_{\mathrm{s}}}\right)\left(N_\omega + \frac{1}{2} \mp \frac{1}{2}\right)\sum_{m,k_x'}\left[\frac{g_{m,k_x'}^{\mathrm{1D}}\left(E \pm \hbar\omega_{ph}\right)}{q_{\mathrm{x}}{}^2}\left(1 - 2I_1\left(\left|q_{\mathrm{x}}R_0\right|\right)K_1\left(\left|q_{\mathrm{x}}R_0\right|\right)\right)\right] \tag{31}$$

Here the sum still remains and refers to the summation of all final states in all available momenta and subbands, and cannot be simplified further from the equation as POP is an anisotropic mechanism, and the $q_{\mathrm{x}}$ still remains within the summation.

Note that using constant wavefunctions as in Ref. [51] is what allowed us to derive simpler semi-analytical expressions for the polar optical phonon scattering based on Bessel functions. It is also a much cheaper computation way rather than employing the actual wavefunction coefficients from tight-binding. As the larger diameter NWs we consider (up to 40 nm) contain ~30,000 atoms, each atom is described by 10 orbitals, the amount of memory required to store all wavefunction coefficients for all states in order to perform the scattering operations will be prohibitive. For strictly cosine/sine-like wavefunctions and infinite barriers, the form factors can be shown to be $9/4A$ for intra-band and $1/A$ for inter-band scattering [19, 20, 42], and in the presence of a large number of subbands the $1/A$ part dominates [44].

For surface roughness scattering (SRS), we use a simplified approach described in our prior works [19], according to which the transition rate is determined by the differential shift in the band edges of the NW upon diameter scaling $\frac{\Delta E_0}{\Delta d}$ [57, 58] as:

$$S_{n,m}^{SRS}\left(k_x, k_x'\right) = \frac{2\pi}{\hbar}\left(\frac{q_0 \Delta E_{\mathrm{C}}}{\Delta d}\right)^2\left(\frac{2\sqrt{2}\Delta_{\mathrm{rms}}{}^2 L_{\mathrm{C}}}{2 + q_x{}^2 L_{\mathrm{C}}{}^2}\right)\delta\left(E_m\left(k_x'\right) - E_n\left(k_x\right)\right), \tag{32}$$

where $q_x = k_x - k_x'$, $E_{\mathrm{C}}$ is the conduction band edge, $d$ is the nanowire diameter, $\Delta_{\mathrm{rms}}$ is the average surface roughness and $L_{\mathrm{C}}$ is the roughness correlation length. We have chosen $\Delta_{\mathrm{rms}} = 1\ \mathrm{nm}$ and $L_{\mathrm{C}} = 2\ \mathrm{nm}$ as these are similar to commonly encountered lengths in experiments [31, 59, 60]. The band edge variation is the dominant SRS mechanism in ultra-scaled channels and results in the low-field mobility in ultra-thin nanostructures to follow a $d^6$ behavior. The scattering rate is then evaluated as previously by:

$$\frac{1}{\tau_n(k_x)} = \frac{2\pi}{\hbar}\left(\frac{e_0 \Delta E_\mathrm{C}}{\Delta d}\right)^2 \frac{1}{L_\mathrm{x}} \sum_{m,k_x'} \left(\frac{2\sqrt{2}\Delta_\mathrm{rms}{}^2 L_\mathrm{C}}{2 + q_x{}^2 L_\mathrm{C}{}^2}\right) g_{m,k_x'}^{1\mathrm{D}}(E) \tag{33}$$

where the energy of the final and initial states are equal ($E' = E$) since SRS is an elastic process.

Calibration to bulk mobility: We begin by computing the phonon-limited low-field mobility for the InAs nanowires as a function of the diameter from $d$ = 3 nm to $d$ = 40 nm. The result is shown in Fig. 3, where the larger diameter NW mobility is ~50,000 $cm^2$/V-s with a slight downward trend with increasing diameter. The bulk phonon-limited low-field mobility value is ~40,000 $cm^2$/V-s [49], and our quantitative overestimation could be that indeed larger NW diameters are needed to reach the bulk mobility, or the deformation potentials chosen, which are bulk values, are not that accurate for NWs. Nevertheless, we still use bulk values, although it is observed that phonon confinement can lead to larger deformation potential values. Our goal is not to accurately map the bulk mobility, but to quantitatively present the trend of the TE coefficients with diameter.

The phonon-limited mobility in Fig. 3 (blue line) is dominated by POP as InAs is a polar material. A slight increase in mobility around $d$ ~ 20 nm arises from the increase in the average $q_\mathrm{x}$ as the number of bands are reduced (see Fig. 1), i.e. the momentum exchange vector that determines the anisotropic behavior of the POP in Eq. 31. For smaller NW diameters, the mobility tends to drastically decrease because of the increase in the phonon form factor [42]. In the case where SRS is included in the simulation (red line in Fig. 3), an even larger reduction in the electron mobility is observed, which becomes more severe as the diameter is reduced.

# III. Results and Discussion

Thermoelectric performance of InAs nanowires: Here we proceed to analyze the behavior of the thermoelectric properties for different NW diameters, first in the case of phonon-limited transport conditions. The conductivity $\sigma$, Seebeck coefficient $S$, and power factor (PF) $\sigma S^2$ versus carrier density for [100] nanowires of diameters from $d$ = 40 nm down to $d$ = 3 nm are shown in Fig. 4a, 4b, and 4c, respectively at $T$ = 300 K. Following the mobility trend, the electrical conductivity for the narrower nanowires is significantly lower compared to that of the larger nanowire diameters (Fig. 4a), with the exception of the $d$ = 10 nm NW (red line), which overpasses all others from densities $n > 10^{18}/cm^3$ and above. This is a consequence of the reduction in the POP scattering rates as the average exchange vector decreases with reduced diameter and reduced number of bands. On the other hand, as the diameter is decreased, a significant increase is observed in the Seebeck coefficient across all carrier concentrations as shown in Fig. 4b. This is a consequence of the increase in the $\eta_F$ as indicated in the inset of Fig. 1d, which essentially increases the average energy of the current flow and consequently the Seebeck coefficient. As a consequence of these trends, the power factor in Fig. 4c exhibits a somewhat erratic behavior, where the narrower nanowires ($d$ < 5 nm) indicate a clear advantage only at higher carrier concentrations, beyond $n = 10^{18}$ / $cm^3$. The power factor is maximized for the $d$ = 10 nm NW at $n = 10^{18}$ / $cm^3$ and for the $d$ = 5 nm NW at $n = 10^{19}$ / $cm^3$. For these wires, the PF reaches large values of > 5 $mW/mK^2$, which signals promising TE performance. The larger diameter NWs ($d$ = 20 and 40 nm), lack significantly in performance, and their peak appears at lower densities.

In Fig. 5 we plot the same quantities, but now we include SRS in the calculations. Now the situation changes in favor of the larger diameter NWs. The electronic conductivity (Fig. 5a) of the smaller diameter NWs is reduced as it suffers significantly from SRS. The Seebeck coefficient, on the other hand, remains very similar to that of the phonon-limited case, where the smaller diameter NWs have higher Seebeck coefficients at the same carrier densities. This is because at first order the Seebeck coefficient is determined by the average energy of the current flow, and does not depend strongly on scattering. The PF in this case, is favored by larger diameters, which have the highest electronic conductivity. The $d$ = 10 nm NW is the one with the middle value of the conductivity and Seebeck coefficient, and it turns out that it also has comparable PF performance to the $d$ = 20 and 40 nm NWs, around 1 $mW/mK^2$.

The $d$ = 10 nm NW can be technologically more challenging to achieve, but it will also have the advantage of lower thermal conductivity. Thus, is Fig. 6 we show an illustration for the TE performance upon diameter scaling for NWs at the same constant carrier concentration of $n = 10^{18}/cm^3$, the density for which the $d$ = 10 nm NW PF peaks.

The TE coefficients $\sigma$, $S$, and $\sigma S^2$ are plotted versus the nanowire diameter, $d$. In this case, we plot the phonon-limited TE coefficients in blue lines, and we then include SRS in addition in red lines. In the phonon-limited transport case, the electrical conductivity in Fig. 6a increases by ~40% as the diameter is scaled down to $d$ ~ 12 nm compared to the larger diameter value, but further diameter scaling results in its sharp drop. This is a consequence of the increase in electron-phonon scattering (form factors) and effective mass increase. As shown in Fig. 2b, for diameters below $d$ = 10 nm and carrier densities $n$ = $10^{18}/cm^3$, only one subband participates in transport and the Fermi level is pushed below the band edge, in which case carriers with lower velocities participate in transport, and the conductivity is reduced. On the other hand, the shift in $E_F$ increases the Seebeck coefficient significantly (Fig. 6b).

Interestingly, the power factor (blue lines in Fig. 6c) experiences a large increase of ~6× compared to the bulk value with diameter scaling, with a peak observed at around $d$ ~ 8 nm. For ultra-narrow diameters the power factor is strongly reduced, dominated by the conductivity reduction. Note that this is a much larger increase compared to what is observed for simulations of materials of heavier effective masses such as in n-type Si, in which case increases are not observed [19, 23]. This is because, in light mass materials subband quantization is stronger, which shifts the $E_F$ much more with confinement, resulting in a much larger increase in $S$. If this effect begins at larger diameters as in InAs, then there is larger room for scaling and larger Seebeck coefficient increases can be achieved as well.

Once SRS is also included in the calculation (red lines in Fig. 6), the increase in conductivity down to $d$ = 10 nm is weakened, and afterwards the conductivity trend is downward with diameter scaling. On the other hand, the Seebeck coefficient in Fig. 6b retains its increasing trend with diameter reduction, with a slight increase over the phonon-limited values. Due to the different behavior of the conductivity, the power factor in Fig. 6c is qualitatively different compared to the phonon-limited trend. Although the strong increase is now absent, still an increase of almost ~2× is observed around $d$ = 12 nm. We would like to stress, however, that the trend in Fig. 6a and 6c depends on the choice of the chosen density, whereas Fig. 4 and Fig. 5 are the ones providing the complete trends. The density chosen is what maximizes the performance of the $d$ = 10 nm NW, which is the one of large enough diameter for significant power factor, but as narrow as possible for ultra-low thermal conductivity.

Our results indicate that power factor benefits in low-dimensional InAs nanowires can be achieved under phonon scattering-limited transport. SRS, however, suppresses these benefits significantly. Improvements in the power factor are a result of: i) large improvements in the Seebeck coefficient after an increase in $\eta_F$ upon confinement, which in turn increases the energy of the current flow, ii) but also because quantization reduces the strength of POP scattering around NW diameters of $d$ = 10 nm. In light mass materials this effect begins at larger diameters, which allows for design flexibility by scaling. On the other hand, this same light effective mass that causes strong confinement, also causes a similarly large SRS as a result of larger sensitivity in the band edges of the electronic structure. Thus, the same effect that provides the benefits, also takes most of them away. In comparison, for heavier effective mass materials, such as Si, moderate improvements in the power factor are observed upon confinement, but at smaller NW diameters, of $d$ ~ 5 nm. At such narrow diameters SRS is also strong, and benefits are also suppressed, even eliminated [23]. Thus, the benefits in polar, light mass materials are expected to be larger compared to non-polar materials with larger effective masses.

Despite the difficulty in achieving power factor improvements in low-dimensional materials (which could be the reason why experimental evidence has not yet been reported), low-dimensional materials can provide very low thermal conductivities $\kappa_l$, originating from enhanced phonon-boundary scattering [6, 7]. The fact that SRS also drastically affects phonons even at a larger degree compared to electrons, makes it so that rough boundaries are actually favorable. However, the knowledge at which length scales and for which materials the power factor is less affected, or even increased, can provide opportunities of improving the *ZT* figure of merit of low-dimensional TE materials.

With regards to comparisons to experiments, out of the several experimental works on InAs nanowires, we have identified two works which provide room temperature measurements for the TE coefficients of NWs with diameter $d$ ~ 20 nm, for which we can perform some comparison between theory and simulation. Direct comparison between theory and simulation is not straight-forward because of the uncertainties in diameter, surface roughness amplitude, and mostly the carrier concentration, upon which the TE coefficients vary significantly. The best way to compare simulation and experiment is to compared at a similar Seebeck coefficient, which is less susceptible to the details of scattering. In Ref. [31], the authors measured the TE properties of a $d$ ~ 20 nm NWs, by using a gate to tune them through tuning the carrier density. At the measured Seebeck coefficient values of -0.2 to -0.12 mV/K, the PF was measured to vary from 1.7 to 1.4 mW/mK$^2$. From Fig. 4, the simulated phonon-limited (upper limit) PF at those Seebeck coefficient values is ~ 3 mW/mK$^2$ (green line). When SRS is introduced in Fig. 5, the PF drops to ~ 1 mW/mK$^2$ for the roughness amplitude of $\Delta_{rms}$ = 1 nm we used, suggesting that the experimental $\Delta_{rms}$ might have been somewhat smaller. In the second work, Ref. [32], the authors measured the TE PF of a $d$ ~ 23 nm InAs NW, again using gating techniques, and found it to be ~0.05 mW/mK$^2$ at densities of $10^{18}$/cm$^3$, which is however significantly lower compared to what we compute, possibly due to numerous other scattering mechanisms present and not accounted for in the simulation.

The $ZT$ figure of merit includes the thermal conductivity, and the overall thermal conductivity is given by the addition of the electronic and the phonon part of the thermal conductivities as $\kappa = \kappa_e + \kappa_l$. The $\kappa_e$ is given by $\kappa_e = L\sigma T$, where $L$ is the Lorenz number. Under the simple acoustic phonon scattering conditions and parabolic bands, the Lorenz number resides mostly between $L = 2.45\times10^{-8}$ W Ω K$^{-2}$ in the degenerate limit and $L = 1.49\times10^{-8}$ W Ω K$^{-2}$ in the non-degenerate limit. These values are routinely used to estimate $\kappa_e$ when limited knowledge about thermal transport details exists. However, we have shown that the Lorenz number can be reduced significantly from the degenerate limit in the presence of multi-band effects, and inter-band scattering [61]. The Lorenz number of the InAs NWs we consider is shown in Fig. 7 for the case of phonon-limited transport (blue line) and phonon plus SRS limited transport (red line) for NWs with a carrier density of $n = 10^{18}$/cm$^3$. Indeed, the Lorenz number at large diameters resides at values around the degenerate limit, as expected since $E_F$ resides well into the bands (Fig. 1d). The Lorenz number takes a sudden drop to the non-degenerate limit (and even below) at $d$ ~12 nm in the presence of SRS, which lowers $\kappa_e$. This is a consequence of the $E_F$ shifting lower, towards non-degenerate conditions, still at the same density. The important thing here, however, is that the power factor can increase (at least at the best case around the $d$ ~ 10 nm NW), and the *ZT* would also benefit from reduction in both $\kappa_l$ and $\kappa_e$. For example, the thermal conductivity of such narrow NWs is reported to be around ~2 mW/mK[2], in which case a *ZT* of ~0.15 can be reached, which is a significant value for room temperature operation.

Finally, we need to elaborate on the assumptions and approximations we have made in this work. We have used an atomistic approach to extract the bandstructure of the NWs, however we considered only pristine, hydrogen passivated NWs, and ignored any strain effects, surface relaxation effects, or defects that could reside in the NW core or surface. In addition, we have assumed bulk phonons when calculating electron-phonon scattering, and ignored any phonon confinement effects, or change in the deformation potential parameters as the diameter is reduced. Finally, the SRS strength is simply determined by the shift in the band edges with diameter change, and we ignored other elements that could contribute to SRS [62]. However, we believe that these will only result in small quantitative changes to our results, and not qualitatively change our conclusions. The method developed, combines atomistic bandstructures with energy-dependent scattering rates based on deformation potential theory within the BTE and is also applied for polar materials. It can be an intermediate between the computationally cheap constant RTA method, and the computationally prohibitive methods which compute scattering rates based on first principles. In addition, the described method can include with relative ease other scattering mechanisms beyond phonons, such as SRS and potentially ionized impurity scattering, something which large codes employed by the TE community do not offer easily.

# IV. Conclusions

In this work, using atomistic full-band electronic structures coupled to the Boltzmann transport method, we theoretically investigated the thermoelectric properties of InAs nanowires with diameters from $d$ = 40 nm down to $d$ = 3 nm. We employ deformation potential theory and energy dependent scattering times, and include the effect of electron-phonon and surface roughness scattering. Under phonon-limited transport conditions, we find that a very large improvement of the power factor of the order of 6× can be potentially achieved as the diameter is scaled to $d$ ~ 10 nm. This is a consequence of improvements in the conductivity due to weakening of the POP scattering rate and an improvement in the Seebeck coefficient. Under surface roughness scattering transport conditions, still an improvement in the power factor of ~2× can be retained at diameters of around $d$ ~ 10 nm. At even narrower diameters, the power factor drops sharply under any scattering consideration due to the strong increase in electron-phonon scattering and surface roughness scattering. The fact that bulk-like, or higher, power factors can be achieved even for diameters as low as $d$ ~ 7 nm, can be quite important for achieving high $ZT$ values since at those diameters the thermal conductivity is significantly reduced. Finally, the method we employ, allows energy dependent relaxation times (something commonly avoided in thermoelectric material studies), and is still computationally efficient in coupling complex bandstructures with Boltzmann transport. This is something that can also be useful for 2D and 3D material simulator development as well.

*Acknowledgement:* This work has received funding from the European Research Council (ERC) under the European Union's Horizon 2020 research and innovation programme (grant agreement No 678763).

# Appendix:

Derivation of Eq. 19

$$\left|D_{1\mathrm{D}}(q)\right|^2 = \sum_{q_\mathrm{R}} \left|I_{k_x',k_x}^{m,n}\right|^2 \left|K_{\vec{q}}\right|^2 \left|A_{\vec{q}}\right|^2$$

$$= \frac{A}{4\pi^2} \int_{\vec{q}_\mathrm{R}} d^2\vec{q}_\mathrm{R} \left[ \int_R d^2R \, e^{\pm i\vec{q}_\mathrm{R}\cdot\vec{R}} \, \frac{F_{m,k_x'}(\vec{R})^*}{\sqrt{A}} \frac{F_{n,k_x}(\vec{R})}{\sqrt{A}} \right]^2 \left|K_{\vec{q}}\right|^2 \left|A_{\vec{q}}\right|^2$$

$$= \frac{A}{4\pi^2} \int_{\vec{q}_\mathrm{R}} d^2\vec{q}_\mathrm{R} \left[ \int_R d^2R \, e^{\pm i\vec{q}_\mathrm{R}\cdot\vec{R}} \, \frac{\Theta(R_0 - R)}{\pi R_0^{\,2}} \right]^2 \left|K_{\vec{q}}\right|^2 \left|A_{\vec{q}}\right|^2$$

$$= \frac{A}{4\pi^2} \int_{\vec{q}_\mathrm{R}} d^2\vec{q}_\mathrm{R} \left[ \frac{1}{\pi R_0^{\,2}} \int_R d^2R \, e^{\pm i\vec{q}_\mathrm{R}\cdot\vec{R}} \right]^2 \left|K_{\vec{q}}\right|^2 \left|A_{\vec{q}}\right|^2$$

$$= \frac{A}{4\pi^2} \int_{\vec{q}_\mathrm{R}} d^2\vec{q}_\mathrm{R} \left[ 2\frac{J_1(q_\mathrm{R}R_0)}{q_\mathrm{R}R_0} \right]^2 \left|K_{\vec{q}}\right|^2 \left|A_{\vec{q}}\right|^2$$

$$= \frac{A}{\pi^2} \int_0^{2\pi}\int_0^{\infty} q_\mathrm{R}\, dq_\mathrm{R}\, d\vartheta \; \frac{J_1^{\,2}(q_\mathrm{R}R_0)}{(q_\mathrm{R}R_0)^2} \left|K_{\vec{q}}\right|^2 \left|A_{\vec{q}}\right|^2$$

$$= \frac{A}{\pi^2} \left[ 2\pi \int_0^{\infty} dq_\mathrm{R} \; \frac{J_1^{\,2}(q_\mathrm{R}R_0)}{q_\mathrm{R}R_0^{\,2}} \left|K_{\vec{q}}\right|^2 \left|A_{\vec{q}}\right|^2 \right]$$

$$= \frac{2A}{\pi} \left[ \int_0^{\infty} dq_\mathrm{R} \; \frac{J_1^{\,2}(q_\mathrm{R}R_0)}{q_\mathrm{R}R_0^{\,2}} \left|K_{\vec{q}}\right|^2 \left|A_{\vec{q}}\right|^2 \right]$$

Derivation of Eq. 20 for ADP scattering:

$$\left|D_{\mathrm{1D}}(q)\right|^2 = \frac{2A}{\pi}\left[\int_0^\infty dq_{\mathrm{R}}\, \frac{J_1^{\,2}(q_{\mathrm{R}}R_0)}{q_{\mathrm{R}}R_0^{\,2}}\, q^2 D_{\mathrm{ADP}}^{\;2}\, \frac{1}{\Omega}\, \frac{\hbar\left(N_\omega + \frac{1}{2} \mp \frac{1}{2}\right)}{2\rho\omega_{\mathrm{ph}}}\right]$$

$$= \frac{2AD_{\mathrm{ADP}}^{\;2}}{\pi}\frac{1}{\Omega}\left[\int_0^\infty d\left(R_0 q_{\mathrm{R}}\right) \frac{J_1^{\,2}(q_{\mathrm{R}}R_0)}{q_{\mathrm{R}}R_0^{\,3}} q^2\, \frac{\hbar\left(2N_\omega\right)}{2\rho\omega_{\mathrm{ph}}}\right]$$

$$= \frac{2AD_{\mathrm{ADP}}^{\;2}}{\pi\rho}\frac{1}{\Omega}\left[\int_0^\infty d\left(R_0 q_{\mathrm{R}}\right) \frac{J_1^{\,2}(q_{\mathrm{R}}R_0)}{q_{\mathrm{R}}R_0^{\,3}} q^2\, \frac{\hbar N_\omega}{\omega_{\mathrm{ph}}}\right]$$

$$= \frac{2AD_{\mathrm{ADP}}^{\;2}}{\pi\rho}\frac{1}{\Omega}\left[\int_0^\infty d\left(R_0 q_{\mathrm{R}}\right) \frac{J_1^{\,2}(q_{\mathrm{R}}R_0)}{q_{\mathrm{R}}R_0^{\,3}} q^2\, \frac{\hbar k_{\mathrm{B}}T}{\hbar\omega_{\mathrm{ph}}^{\;2}}\right]$$

$$= \frac{2AD_{\mathrm{ADP}}^{\;2}}{\pi R_0^{\,2}\rho}\frac{1}{\Omega}\left[\int_0^\infty d\left(R_0 q_{\mathrm{R}}\right) \frac{J_1^{\,2}(q_{\mathrm{R}}R_0)}{q_{\mathrm{R}}R_0} q^2\, \frac{k_{\mathrm{B}}T}{v_{\mathrm{s}}^{\,2}q^2}\right]$$

$$= \frac{2AD_{\mathrm{ADP}}^{\;2}k_{\mathrm{B}}T}{A\rho v_{\mathrm{s}}^{\,2}}\frac{1}{\Omega}\left[\int_0^\infty d\left(R_0 q_{\mathrm{R}}\right) \frac{J_1^{\,2}(q_{\mathrm{R}}R_0)}{q_{\mathrm{R}}R_0}\right]$$

$$= \frac{2D_{\mathrm{ADP}}^{\;2}k_{\mathrm{B}}T}{\rho v_{\mathrm{s}}^{\,2}}\frac{1}{\Omega}\left[\frac{1}{2}\right]$$

$$= \frac{1}{\Omega}\frac{D_{\mathrm{ADP}}^{\;2}k_{\mathrm{B}}T}{\rho v_{\mathrm{s}}^{\,2}}$$

Derivation of Eq. 24 for ODP scattering:

$$\left|D_{\text{1D}}\left(q\right)\right|^2 = \sum_{q_\text{R}} \left|I_{k_x',k_x}^{m,n}\right|^2 \left|K_{\vec{q}}\right|^2 \left|A_{\vec{q}}\right|^2$$

$$= \frac{2A}{\pi}\left[\int_0^\infty dq_\text{R} \, \frac{J_1^{\,2}\left(q_\text{R} R_0\right)}{q_\text{R} R_0^{\,2}} D_\text{ODP}^{\,2} \frac{1}{\Omega} \frac{\hbar\left(N_\omega + \frac{1}{2} \mp \frac{1}{2}\right)}{2\rho\omega_\text{ph}}\right]$$

$$= \frac{2A}{\pi} D_\text{ODP}^{\,2} \frac{1}{\Omega} \frac{\hbar\left(N_\omega + \frac{1}{2} \mp \frac{1}{2}\right)}{2\rho\omega_\text{ph}} \left[\int_0^\infty dR_0 q_\text{R} \, \frac{J_1^{\,2}\left(q_\text{R} R_0\right)}{q_\text{R} R_0^{\,3}}\right]$$

$$= \frac{2A}{\pi R_0^{\,2}} D_\text{ODP}^{\,2} \frac{1}{\Omega} \frac{\hbar\left(N_\omega + \frac{1}{2} \mp \frac{1}{2}\right)}{2\rho\omega_\text{ph}} \left[\int_0^\infty d\left(R_0 q_\text{R}\right) \frac{J_1^{\,2}\left(q_\text{R} R_0\right)}{q_\text{R} R_0}\right]$$

$$= \frac{2A}{A} D_\text{ODP}^{\,2} \frac{1}{\Omega} \frac{\hbar\left(N_\omega + \frac{1}{2} \mp \frac{1}{2}\right)}{2\rho\omega_\text{ph}} \left[\frac{1}{2}\right]$$

$$= \frac{1}{\Omega} D_\text{ODP}^{\,2} \frac{\hbar\left(N_\omega + \frac{1}{2} \mp \frac{1}{2}\right)}{2\rho\omega_\text{ph}}$$

Derivation of Eq. 28 for POP scattering:

$$\left|D_{\mathrm{1D}}\left(q\right)\right|^2 = \sum_{q_{\mathrm{R}}}\left|I_{k_x',k_x}^{m,n}\right|^2\left|K_{\vec{q}}\right|^2\left|A_{\vec{q}}\right|^2$$

$$= \frac{2A}{\pi}\left|A_{\vec{q}}\right|^2\left[\int_0^\infty dq_{\mathrm{R}}\,\frac{J_1^{\,2}\left(q_{\mathrm{R}}R_0\right)}{q_{\mathrm{R}}R_0^{\,2}}\frac{\rho e^2\omega_{\mathrm{ph}}^{\,2}}{\left(q_{\mathrm{R}}^{\,2}+q_{\mathrm{x}}^{\,2}\right)}\left(\frac{1}{\varepsilon_\infty}-\frac{1}{\varepsilon_{\mathrm{s}}}\right)\right]$$

$$= \frac{2A}{\pi}\rho e_0^{\,2}\omega_{\mathrm{ph}}^{\,2}\left(\frac{1}{\varepsilon_\infty}-\frac{1}{\varepsilon_{\mathrm{s}}}\right)\left|A_{\vec{q}}\right|^2\left[\int_0^\infty dq_{\mathrm{R}}\,\frac{J_1^{\,2}\left(q_{\mathrm{R}}R_0\right)}{q_{\mathrm{R}}R_0^{\,2}}\frac{1}{\left(q_{\mathrm{R}}^{\,2}+q_{\mathrm{x}}^{\,2}\right)}\right]$$

$$= \frac{2A}{\pi}\rho e_0^{\,2}\omega_{\mathrm{ph}}^{\,2}\left(\frac{1}{\varepsilon_\infty}-\frac{1}{\varepsilon_{\mathrm{s}}}\right)\left|A_{\vec{q}}\right|^2\left[\int_0^\infty d\left(R_0q_{\mathrm{R}}\right)\frac{J_1^{\,2}\left(q_{\mathrm{R}}R_0\right)}{q_{\mathrm{R}}R_0}\frac{1}{\left(\left[R_0q_{\mathrm{R}}\right]^2+\left[R_0q_{\mathrm{x}}\right]^2\right)}\right]$$

$$= \frac{2A}{\pi}\rho e_0^{\,2}\omega_{\mathrm{ph}}^{\,2}\left(\frac{1}{\varepsilon_\infty}-\frac{1}{\varepsilon_{\mathrm{s}}}\right)\left|A_{\vec{q}}\right|^2\left[\frac{1}{q_{\mathrm{x}}^{\,2}R_0^{\,2}}\left(\frac{1}{2}-I_1\left(\left|q_{\mathrm{x}}R_0\right|\right)K_1\left(\left|q_{\mathrm{x}}R_0\right|\right)\right)\right]$$

$$= \frac{2A}{\pi R_0^{\,2}}\rho e_0^{\,2}\omega_{\mathrm{ph}}^{\,2}\left(\frac{1}{\varepsilon_\infty}-\frac{1}{\varepsilon_{\mathrm{s}}}\right)\left|A_{\vec{q}}\right|^2\left[\frac{1}{q_{\mathrm{x}}^{\,2}}\left(\frac{1}{2}-I_1\left(\left|q_{\mathrm{x}}R_0\right|\right)K_1\left(\left|q_{\mathrm{x}}R_0\right|\right)\right)\right]$$

$$= \frac{2A}{A}\rho e_0^{\,2}\omega_{\mathrm{ph}}^{\,2}\left(\frac{1}{\varepsilon_\infty}-\frac{1}{\varepsilon_{\mathrm{s}}}\right)\frac{1}{\Omega}\frac{\hbar\left(N_\omega+\frac{1}{2}\mp\frac{1}{2}\right)}{2\rho\omega_{\mathrm{ph}}}\left[\frac{1}{q_{\mathrm{x}}^{\,2}}\left(\frac{1}{2}-I_1\left(\left|q_{\mathrm{x}}R_0\right|\right)K_1\left(\left|q_{\mathrm{x}}R_0\right|\right)\right)\right]$$

$$= e^2\omega_{\mathrm{ph}}\hbar\left(\frac{1}{\varepsilon_\infty}-\frac{1}{\varepsilon_{\mathrm{s}}}\right)\frac{\left(N_\omega+\frac{1}{2}\mp\frac{1}{2}\right)}{\Omega}\left[\frac{1}{q_{\mathrm{x}}^{\,2}}\left(\frac{1}{2}-I_1\left(\left|q_{\mathrm{x}}R_0\right|\right)K_1\left(\left|q_{\mathrm{x}}R_0\right|\right)\right)\right]$$

Figure 1:

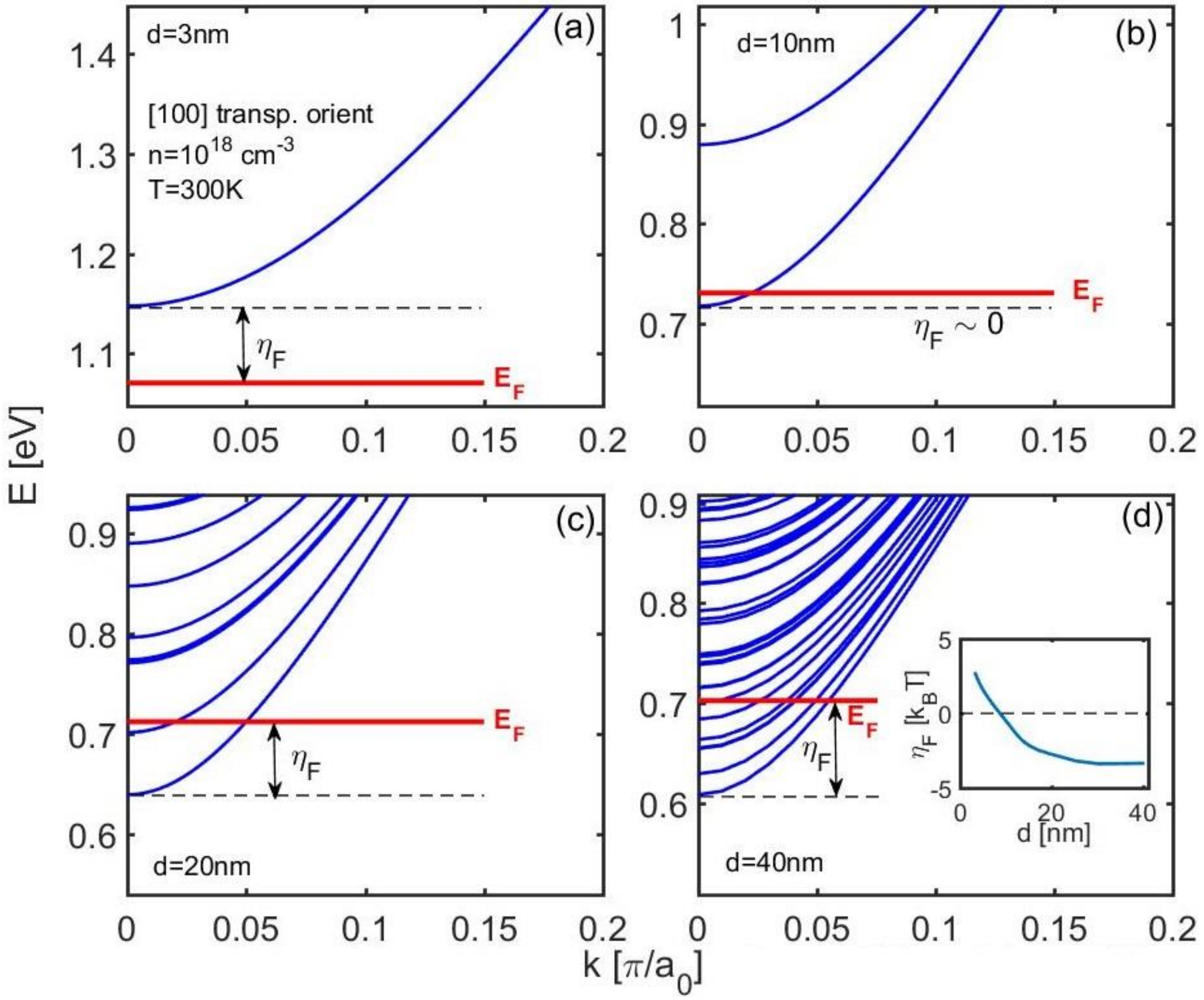

<u>Figure 1 caption:</u>

Electronic bandstructures for [100] InAs nanowires of diameters (a) $d$ = 3 nm, (b) $d$ = 10 nm, (c) $d$ = 20 nm, and (d) $d$ = 40 nm. The position of the Fermi level $E_F$ for carrier density $n = 10^{18}$ / $cm^3$ at room temperature is indicated. The difference of the Fermi level from the band edges $\eta_F = (E_C - E_F)/k_B T$, which determines the Seebeck coefficient is indicated as well. The inset of (d) shows $\eta_F$ versus the NW diameter.

Figure 2:

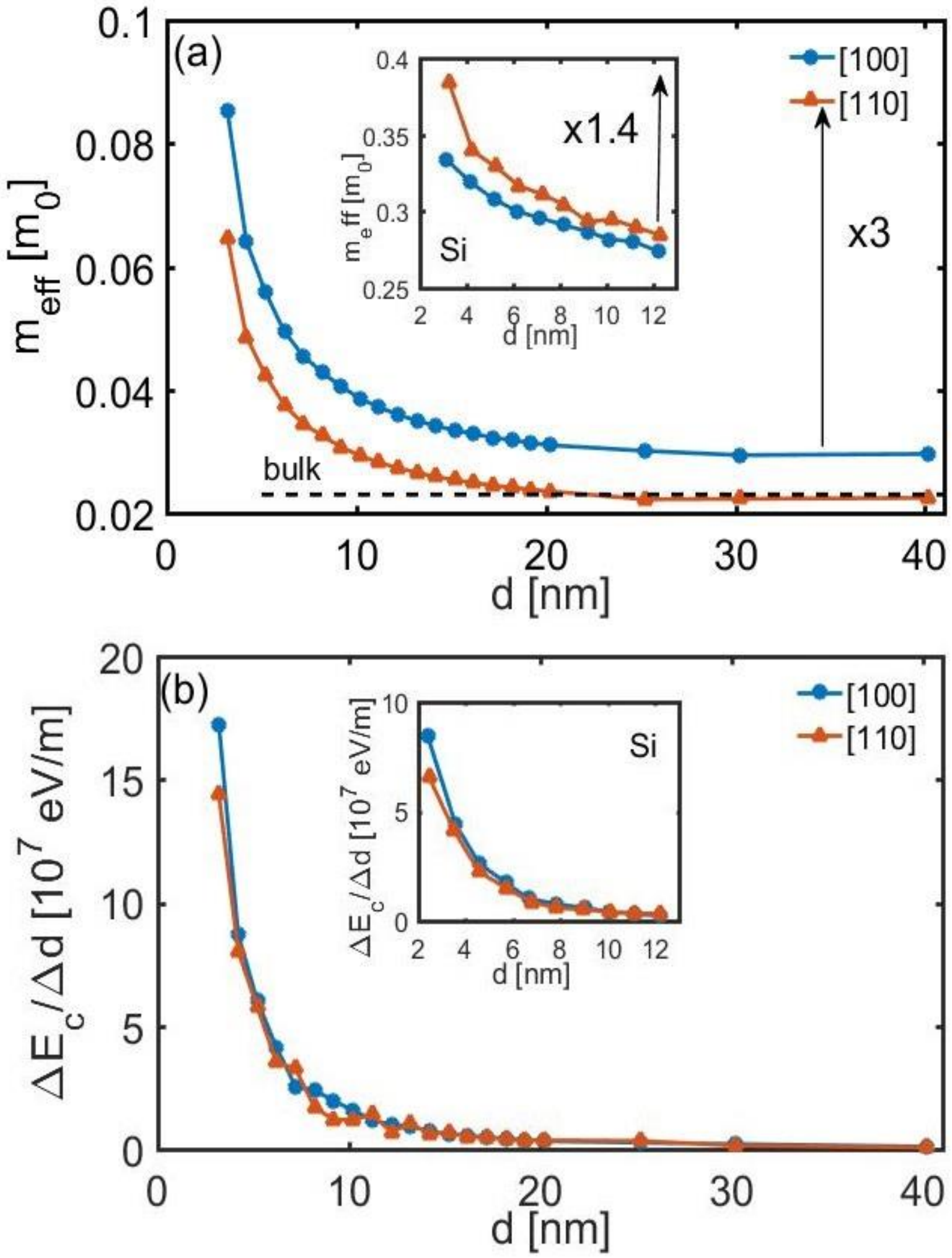

## Figure 2 caption:

(a) The effective mass of the first subband of InAs nanowires as a function of the nanowire diameter. An increase in the mass is observed as the diameter is reduced. (b) The differential of the band edge of the nanowires versus their diameter. Nanowire orientations in [100] (blue-circle lines) and in [110] (orange-triangle lines) are shown. The insets show the corresponding mass variation and band edge differential changes for Si nanowires, as shown in Refs [42, 44], which indicate less variation for both quantities.

Figure 3:

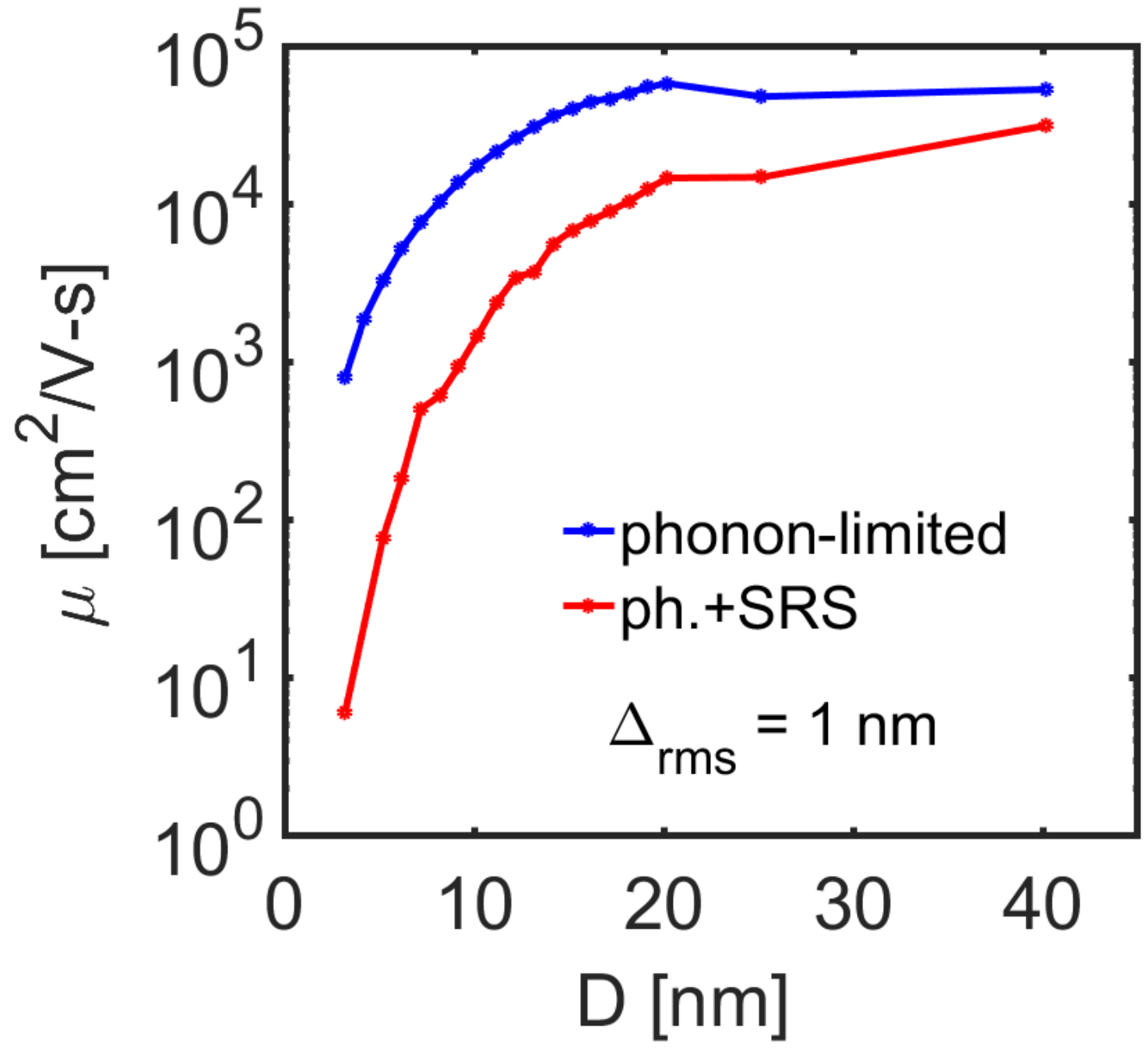

<u>Figure 3 caption:</u>

Low-field electron mobility vs. nanowire diameter for [100] InAs nanowires at room temperature. Different scattering cases are shown: i) the blue line shows phonon-limited transport (including acoustic and optical deformation potentials scattering and polar optical phonon scattering), ii) the red line shows the case when surface roughness scattering (SRS) is added in addition to phonon scattering.

Figure 4:

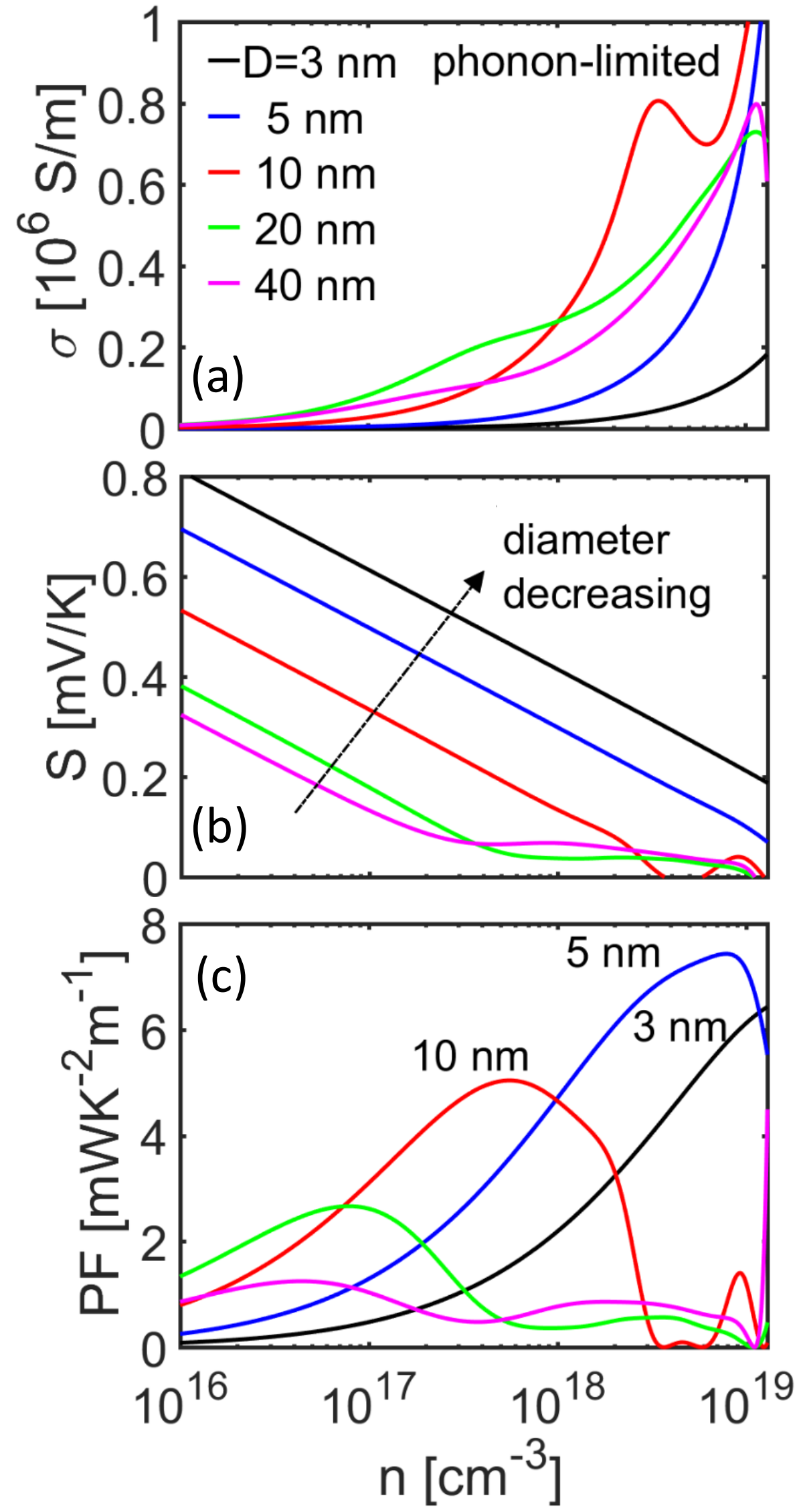

Figure 4 caption:

Thermoelectric coefficients under phonon scattering-limited transport conditions at room temperature for [100] InAs with different diameters, as indicated in the figure. (a) Electrical conductivity, (b) Seebeck coefficient, and (c) power factor versus carrier concentration. As the diameter is reduced, the Seebeck coefficient is increased. The power factor is increased for the smaller diameters around $d$ ~ 3-10 nm.

Figure 5:

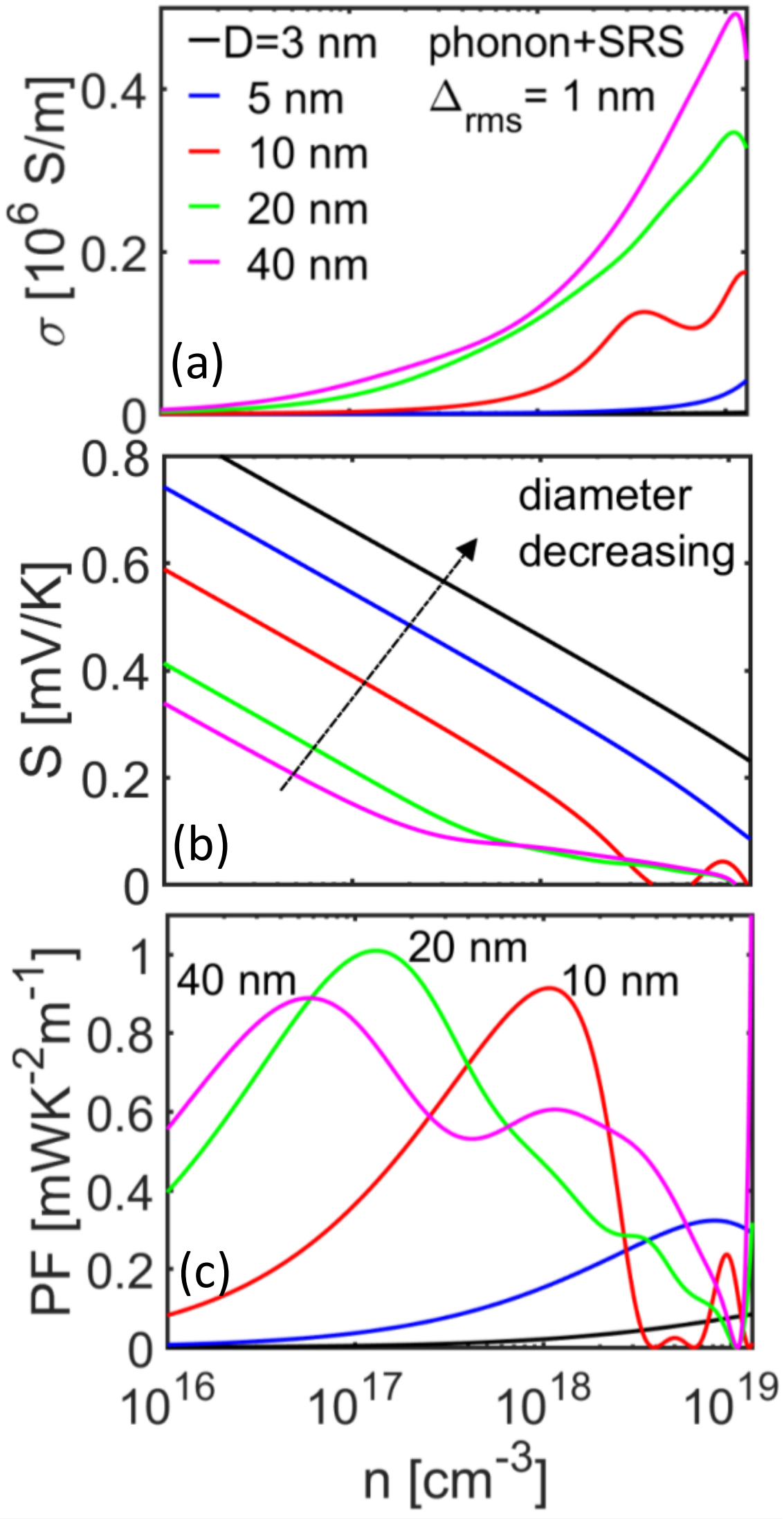

## <u>Figure 5 caption:</u>

Thermoelectric coefficients under phonon plus surface roughness scattering (SRS) transport conditions at room temperature for [100] InAs with different diameters, as indicated in the figure. (a) Electrical conductivity, (b) Seebeck coefficient, and (c) power factor versus carrier concentration. As the diameter is reduced, the Seebeck coefficient is increased. The power factor is increased for the larger diameters around $d$ ~ 10-40 nm.

Figure 6:

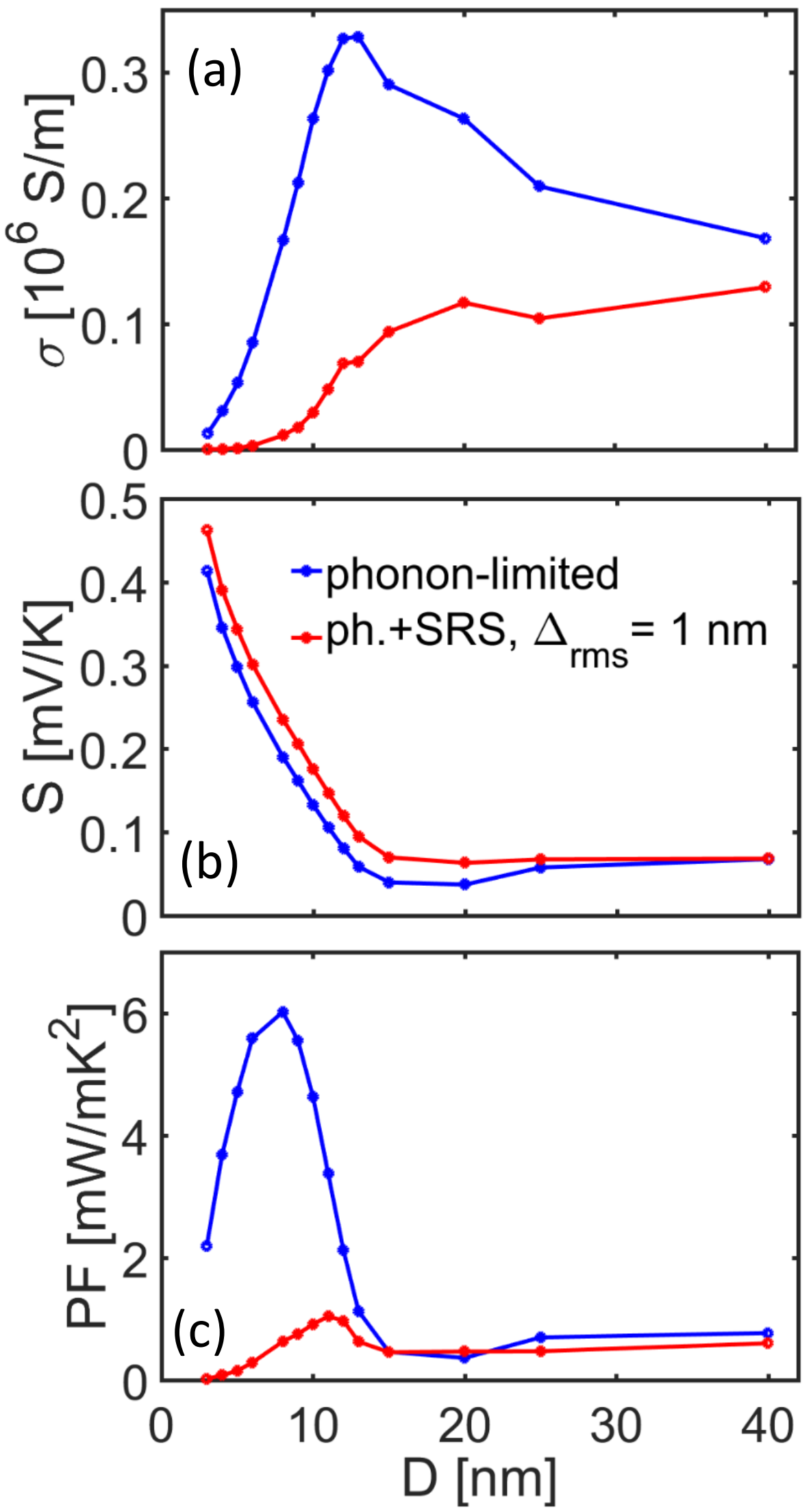

<u>Figure 6 caption:</u>

Thermoelectric coefficients under phonon (blue lines) and phonon plus surface roughness scattering (red lines) transport conditions at room temperature for [100] InAs NWs versus diameter at a fixed carrier concentration of $n = 10^{18}/\text{cm}^3$. (a) Electrical conductivity, (b) Seebeck coefficient, and (c) power factor. As the diameter is reduced, the Seebeck coefficient is increased. The power factor is increased for diameters from around $d \sim 10$ nm.

Figure 7:

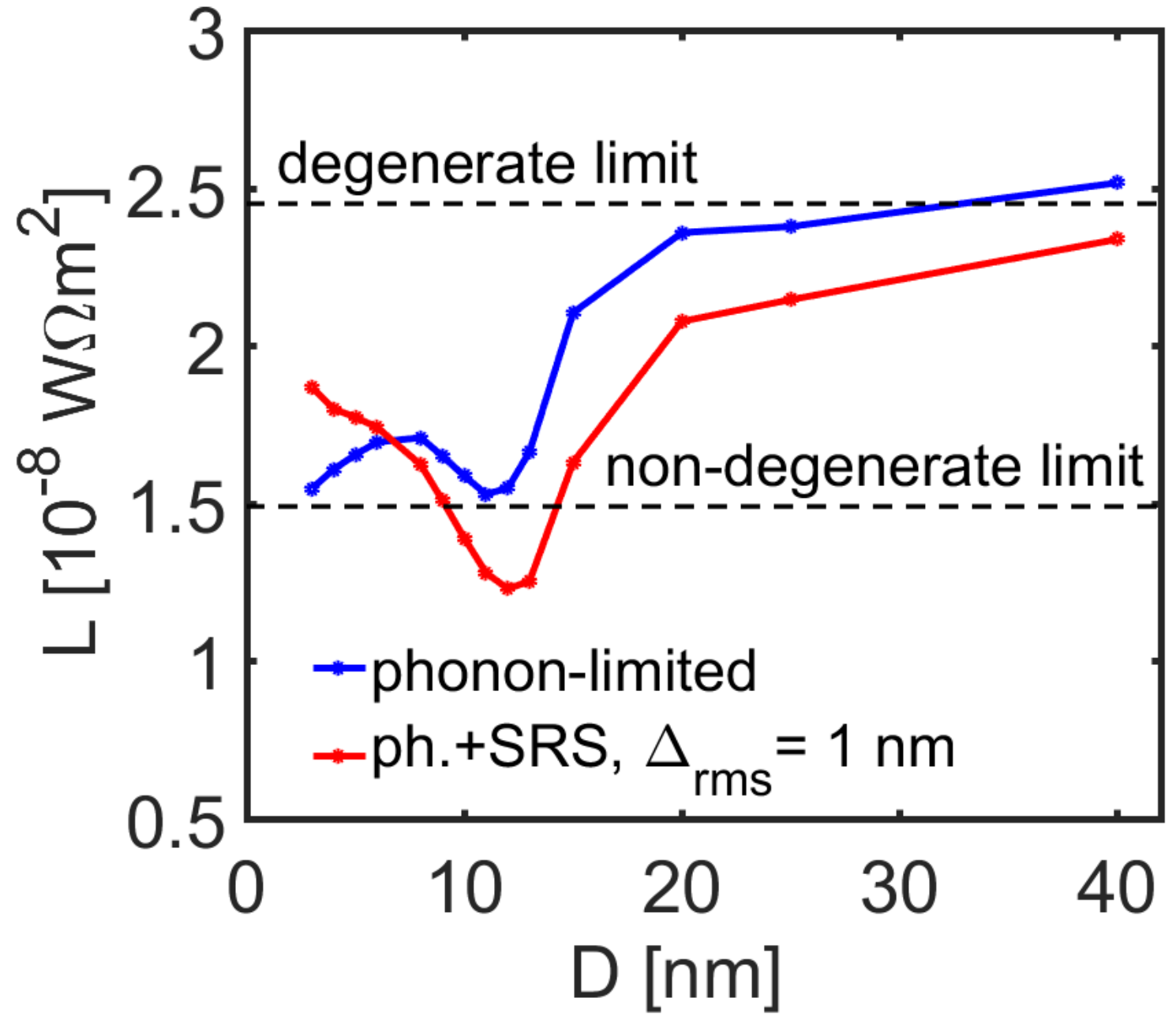

<u>Figure 7 caption:</u>

The Lorenz number versus nanowire diameter under phonon scattering conditions (blue line) and under phonon plus surface roughness scattering (red line) transport conditions at room temperature at a fixed carrier concentration of $n = 10^{18}/cm^3$.